\documentclass[preprint,prd,showpacs,aps]{revtex4}

\usepackage{amsmath}
\usepackage{amsbsy}
\usepackage{amsthm}
\usepackage{amssymb}
\usepackage{latexsym}
\usepackage[dvips]{graphicx}

\begin{document}

\title{Constituent quark masses from a modified perturbative QCD} 

\author{Alejandro Cabo Montes de Oca}
\affiliation{Instituto de Cibern\'{e}tica, Matem\'{a}tica y\\
F\'{\i}sica; Calle E, No. 309, Vedado, La Habana, Cuba}
\affiliation{International Institute for Theoretical and Applied Physics,\\
UNESCO and Iowa State University, Ames, Iowa 50011, U.S.A. }

\author{Marcos Rigol}
\affiliation{Centro de Estudios Aplicados al Desarrollo Nuclear;\\
Calle 30, N. 502 e/ 5ta y 7ma, Miramar, La Habana, Cuba}

\begin{abstract}

A recently proposed modified perturbative expansion for QCD incorporating
gluon condensation is employed to evaluate the quark and gluon self-energy
corrections in the first approximations. The results predict mass values of
1/3 of the nucleon mass for the light quarks $u$, $d$, and $s$ and a
monotonously growing variation with the current mass. The only
phenomenological input consists in that $\langle G^2\rangle $ is evaluated
up to order $g^2$ as a function of the unique parameter $C$ defining the
modified propagator and then $C$ is fixed to give a current estimate of $
\langle g^2G^2\rangle $. The light quarks $u$ and $d$ result to be confined
and the $s$, $c$, $b$ and $t$ ones show damped propagation modes, then
suggesting a model for the large differences in stability between the
nucleons and the higher resonances. The above properties of quark modes
diverge from the fully confinement result following from the similar gluon
propagator previously considered by Munczek and Nemirovski. On another hand,
the condensate effects on the gluon self-energy furnish a tachionic mass
shell as predicted by the Fukuda analysis of gluon condensation in QCD.

\end{abstract}

\pacs{12.38.-t, 12.38.Bx, 12.40.Yx, 12.38.Aw, 12.15.Ff}

\maketitle

\section{Introduction}

One of the great achievements of theoretical High Energy Physics in the last
thirty years has been the discovering and development of the Quantum
Chromodynamics (QCD). The smallness of the coupling constant at high momenta
(asymptotic freedom) has allowed to develop a perturbative framework
applicable for very high energy processes. That situation strongly
simplifies the study of such phenomena and the calculated quantities are in
good agreement with the experimental data. However, perturbative QCD (PQCD)
is far from being able to furnish even a rough description for the
relevant physics at low energies. The solution of this situation is
currently one of the main challenges of Particle Physics.

A relevant phenomenon related with QCD is the color confinement. Nowadays,
there are strong reasons to believe that the relation between the basic
quantities in QCD, like the gluon and quarks fields, and the real world
characterized by a whole variety of interacting mesons and baryons, can be
understood by solving the confinement problem. The basic picture which seems
plausible to be derived from QCD is that the fundamental fields can not be
associated to asymptotic states of quark and gluons and then the true
physical states consist only of colorless composites of quarks and gluons
(mesons and baryons).

The above quoted limitations of the PQCD mean in particular that the usual
Fock space vacuum of the non-interacting theory, is unable to predict, even
approximately, the real ground state properties of QCD \cite
{Savvidy1,Savvidy2,Savvidy3,Reuter}. This is at contrast with the case in
QED where the standard perturbation theory gives a better than good
concordance with the experimental data. Then, the nature of vacuum structure
is one of the main problems to be clarified and naturally its solution is
closely linked with the color confinement effect. Reviews of various 
models considered to investigate the confinement and the vacuum structure
in QCD can be found in Refs. \cite{ShuryakTex,Shuryak2}.

Another basic problem for QCD is the one related with the quark masses, that
is, the so called current quarks masses which appear in the Lagrangian. As
the quarks are not observed as free states, the meaning of the quark masses
need to be considered with care \cite{Gell,Gross,Weinberg1}. The nature of
these parameters is an important object of research since long time ago 
\cite{Weinberg2,Weinberg3,Weinberg4}. The relations among these masses and 
their magnitude have been determined through the methods of current algebra in
combination with measurements of the light mesons masses.

In the former works \cite{1995,PRD,tesis} an attempt directed to construct a
modified perturbation expansion for QCD being able to predict at least some
low energy properties of this theory have been considered. First, in \cite
{1995}, the modified expansion conserving the color $SU(3)$ and Lorentz
symmetries was proposed, aiming to solve the symmetry limitations of the
earlier chromo-magnetic field models \cite{Savvidy1,Savvidy2,Savvidy3,Reuter}
which inspired the search. A similar lack of manifest Lorentz invariance had
also a modification of the Feynman rules, intending to include gluon
condensation, advanced later in \cite{Hoyer1,Hoyer2,Hoyer3} in which a delta
function for $k<\Lambda _{qcd}$ was summed to the perturbative piece. The
expansion proposed in \cite{1995}, considers a change of the gluon
propagator in a term associated to a condensation of zero momentum gluons. This
alteration of the usual rules had the interesting property of producing a
non vanishing value for the gluon condensation parameter $\langle
g^2G^2\rangle $ already in the first approximations \cite{Zakharov}. In
addition, in \cite{1995} a non vanishing value for the effective self-energy
of the gluons was obtained at the loop level. Finally, a perturbative
evaluation of the effective potential as a function of the condensate
parameter indicated that the condensation is spontaneously generated from a
zero condensate state.

The next work \cite{PRD,tesis}, had as a main objective to justify the
applicability of the Feynman expansion introduced in \cite{1995}. This
search was in need because after modifying the propagator, it was unknown
whether or not the initial wavefunction generating the expansion was a
physical state of the free theory. In \cite{PRD,tesis}, by making use of the
operational formulation of QCD developed by Kugo and Ojima \cite{Kugo}, it
was possible to find a physical state of the non-interacting theory being
able to generate through the Wick expansion, the sort of propagators
considered in \cite{1995}. The discussion allowed also a more precise
characterization of the class of changes admitted in the diagrammatic
expansion by the physical state condition on the initial state.
Specifically, it was found that the $C$ parameter, describing the gluon
condensation, must be real and positive.

The present work is an expanded version of a letter sent to publication. Its
main objective is to evaluate first $\langle G^2\rangle $ in terms of the
condensate parameter $C\ $up to order $g^2$, in order to determine
the quark masses after fixing the parameter $g^2C$ to reproduce the accepted
value for $\langle g^2G^2\rangle$. The essential result is the evaluation
of the constituent masses for the light quarks. In order to perform it, we
have made use of the recent values reported for the current masses \cite
{Report}. The idea inspiring the work was the possibility that the gluon
condensation described by modified expansion could predict the appearance of
large masses for the light quarks ($u$, $d$ and $s$) of the order of 1/3 of
the nucleon mass (constituent quark mass), thus furnishing an explanation
for the successful constituent quark models. Surprisingly, after fixing the
condensate parameter $g^2C$ to reproduce the most accepted value of the 
$\langle g^2G^2\rangle$ (through the here performed evaluation of this
quantity in terms of $g^2C$), it followed that the light quarks $u$, $d$ and 
$s$ got exactly the needed constituent like values for the masses in one of
the propagation modes of the modified mass shell. The free propagation
modes, which are at discrepancy with the absence of free quarks, are here
only arising within an approximation evaluating the main effects of the
condensate. After the inclusion of all the terms completing the one loop
approximation in the gluon self-energy an interesting picture arises. It
follows that the previously free propagation modes of the $u$ and $d$ quarks
turn to be confined. On another hand, the $s$, $c$, $b$ and $t$ quarks waves
become damped. That is, the squared rest mass get complex values. It should
be stressed that the peaked structure of the propagator for the $u$ and $d$
at $\sqrt{p^2}=m=0.30915\ GeV$ signals the value of $m$ as the one expected
to appear in the kinetic energy terms $\overrightarrow{p}^2/(2m)$ defining
the free part of the bound state equations for $u$ and $d$ constituted
nucleons. Henceforth, a possible mechanism contributing to justify the
drastic difference between the stability of nucleons and the higher
resonances is suggested. The absence of propagating modes for the $u$ and $d$, 
might be assuring the absolute stability shown by the nucleons bound
states. In other manner, the presence of damped modes in the rest of the
quarks would contribute to the decaying channels observed for all the hadron
resonances associated to these short lived states.

Next, after considering the ground state within each of the baryon resonance
groups as classified in \cite{Report} and its set of three constituent
quarks, and adding the calculated quark masses for each of them (which
depends on the flavor through the assumed values of the current masses), a
spectrum of baryon masses was estimated. The results again reproduce 
reasonably well the experimentally determined resonance energies \cite{Report}.
This outcome suggests the less relevant role of the interaction energy among
quarks in determining the masses of these hadrons.

At this point it is important to make clear that earlier works \cite
{Munczek,Burden} introduced a pure delta function at zero momentum in the
gluon propagator, searching for a model of meson resonances. In these
treatments the constants introduced, multiplying the delta function, were
different to the one fixed in our previous work \cite{PRD,tesis}, leading 
also to very different results. In particular, the singularity structure of
the quark propagator have no poles on the real $p^2$ axis \cite{Burden} at
difference with our results. This difference is directly connected with the
definite sign for the coefficient of the $\delta $-function which in our
case was precisely determined in \cite{PRD,tesis}. Since our analysis in
that paper was based in the direct construction of the state incorporating
the condensate, it follows that the difference with the results of \cite
{Munczek,Burden} is rooted in a corresponding difference in the initial
state determining the Wick expansion. It could be also possible that both
procedures lead to the same exact physical predictions, but the first
approximations in each approach could turn to be more appropriate for the
discussion of specific physical issues. This question will be addressed
elsewhere. It should be also further noticed that gluon condensate states
reproducing the modified expansion of \cite{Munczek,Burden} have been also
discussed in \cite{celenza,pavel}. However, all these references are related
with states showing confined quarks and massive gluons, at variance with our
proposed state determining massive quarks (in the simplest approximation)
and tachyonic gluons.

It is also following that another dispersion relation is obtained which
predicts a vanishing value of the mass as $m_Q\rightarrow 0$. This mode
could be connected with the family of low mass mesons (e.g. $\Pi $ and $K$
mesons). Specifically, the possibility is open that the bound states of
quark excitations in these light mass states of quarks could describe such
low lying mesons within the here presented framework. A circumstance
pointing in this direction is that the spontaneous chirality breaking
associated with this mass generation should be expected to be associated
with some massless Goldstone modes. Therefore, it is possible that the bound
states of such modes predicted by the Bethe-Salpeter equations turn to
describe the light mesons $\Pi$ and $K$. This question however, will be
deferred to a further study.

The calculations of $\langle G^2\rangle $ and the quark and gluon masses
will be performed up order $g^2$. It seems worth anticipating that the
approximation scheme will resemble a sort of quasi-classical limit in which
the condensate seems to play a role of a macroscopic mean quantum field.
Such an interpretation is in concordance with the initial motivation of our
analysis as directed to construct a covariant version of the chromo-magnetic
models \cite{1995}.

Another conclusion which should be noticed is that the evaluated gluon mass
determined from the allowed values of the condensate parameter $g^2C$ is
tachyonic. The possibility for a tachyonic gluon mass has been recently
discussed in the literature \cite{tachyon1,tachyon2,Hoyer3}. This result
indicates that the free propagator of the expansion is not reproducing
itself after the first approximation. This property, however, does not mean
a limitation of the approach. This is because, at variance with 
\cite{Munczek,Burden}, we are not assuming that the exact propagator has the 
same delta function structure. This form only arose for the free propagator as
determined by the Wick expansion around the considered initial state.

The presentation is organized as follows: Section II is devoted to make
precise the elements characterizing the modified perturbation expansion. In
the Section III the calculation of the $\langle G^2\rangle $ up to $g^2$
order for gluodynamics is presented and the result used to estimate the
condensation parameter $g^2C$. Further in Section IV the effects of the
condensate on the quark masses are evaluated and the obtained values used to
estimate the spectrum of the ground states in each baryon group as
classified in \cite{Report}. In Section V the calculation of the effect of
the condensate on the one loop self-energy for the gluons is evaluated.
Finally, the results are summarized and some open questions and a conjecture
are underlined in the last Section VI.

\section{The Modified Feynman Rules}

In this introductory section, the main elements and conventions of the
modified expansion introduced in \cite{1995,PRD,tesis} will be reviewed. The
generating functional of the Green functions in the modified theory has the
form
\begin{eqnarray}
Z\left[ J,\bar{\eta},\eta,\bar{\xi},\xi \right] &=&\frac 1N\int D\left( A^r,
\bar{c}^r,c^r,\bar{\Psi}^r,\Psi ^r\right) \exp \left\{ i\int d^4x{\cal L}
^{sources}\right\},  \label{funct1} \\
{\cal L}^{sources} &=&{\cal L}+J^{\mu,a}A_\mu ^{ra}+\overline{c}^{ra}\eta
^a+\bar{\eta}^ac^{ra}+{\bar{\Psi}}^{ri}{\xi }^i+{\bar{\xi}}^i{\Psi }^{ri}, \\
N &=&\int D\left( A^r,\bar{c}^r,c^r,\bar{\Psi}^r,\Psi ^r\right) \exp \left\{
i\int d^4x{\cal L}\right\},
\end{eqnarray}
where sources for all the renormalized fields $(A_\mu ^{ra},\overline{c}
^{ra},c^{ra},{\bar{\Psi}}^{ri}{,\Psi }^{ri})$ have been introduced in the
usual manner, and the effective Lagrangian for the bare fields was selected
in the Kugo-Ojima quantization procedure for gauge theory as
\begin{eqnarray}
{\cal L} &=&{\cal L}_G+{\cal L}_{Gh}+{\cal L}_Q \\
{\cal L}_G &=&-\frac 14F_{\mu \nu }^a\left( x\right) F^{\mu \nu,a}\left(
x\right) -\frac 1{2\alpha }\left( \partial ^\mu A_\mu ^a\left( x\right)
\right) ^2, \\
{\cal L}_{Gh} &=&-i\partial ^\mu \overline{c}^a\left( x\right) D_\mu
^{ab}\left( x\right) c^b\left( x\right), \\
{\cal L}_Q &=&\bar{\Psi}^i\left( x\right) \left( i\gamma ^\mu D_\mu
^{ij}-m_Q^{*}\delta ^{ij}\right) \Psi ^j\left( x\right).
\end{eqnarray}

The sum over the six quark flavors will be omitted everywhere in order to
simplify the exposition. We think that no confusion should arise for it. The
gluon field intensity has the usual form
\[
F_{\mu \nu }^a\left( x\right) =\partial _\mu A_\nu ^a\left( x\right)
-\partial _\nu A_\mu ^a\left( x\right) +g^{*}f^{abc}A_\mu ^b\left( x\right)
A_\nu ^c\left( x\right), 
\]
where $D_\mu ^{ab},\ D_\mu ^{ij}$ are the covariant derivatives in the adjoint
and fundamental representations of the group $SU(3)$ respectively
\begin{eqnarray*}
D_\mu ^{ab}\left( x\right) &=&\partial _\mu \delta ^{ab}-g^{*}f^{abc}A_\mu
^c\left( x\right), \\
D_\mu ^{ij}\left( x\right) &=&\partial _\mu \delta ^{ij}-ig^{*}T^{ij,a}A_\mu
^a.
\end{eqnarray*}

Here, the bare fields and bare coupling constant and masses ($g^{*},\ m_Q^{*}$), 
were expressed in terms of their renormalized counterparts
following the same conventions of Ref. \cite{Muta}. Below, the renormalized
masses and coupling will be denoted as $m_Q$ and $g$ respectively.

It should be recalled that as argued in \cite{PRD,tesis} the physical
predictions for the value $\alpha =1$ of the gauge parameter should have
physical meaning whenever the adiabatic connection of the interaction does
not lead the evolving state out the physical space. Therefore here we will
fix this value of the parameter.

Then, after the standard procedure of extracting out of the functional
integral the exponential of the terms of higher than second order in the
fields (vertices) the generating functional takes the form to be employed in
the evaluations below \cite{Faddeev}:
\begin{eqnarray}
&&Z\left[ J,\bar{\eta},\eta,\bar{\xi},\xi \right] =N^{-1}\exp \left\{
i\left[ \frac{S_{abc}^G}{3!i^3}\frac{\delta ^3}{\delta J_a\delta J_b\delta
J_c}+\frac{S_{abcd}^G}{4!i^4}\frac{\delta ^4}{\delta J_a\delta J_b\delta
J_c\delta J_d}\right. \right.  \nonumber \\
&&\qquad\qquad\qquad\qquad +\left. \left. \frac{S_{ras}^{Gh}}{2!i^3}
\frac{\delta ^3}{\delta 
\bar{\eta}_r\delta J_a\delta \left( -\eta _s\right) }+\frac{S_{iaj}^Q}{i^3}
\frac \delta {\delta \bar{\xi}_i\delta J_a\delta \left( -\xi _j\right) }
\right] \right\} Z_0\left[ J,\bar{\eta},\eta,\bar{\xi},\xi \right],
\label{funct2}
\end{eqnarray}
with the free functional given by
\[
Z_0\left[ J,\bar{\eta},\eta,\bar{\xi},\xi \right] =\exp \left\{ i\frac{
J_aG_G^{ab}J_b}2+i\bar{\eta}_rG_{Gh}^{rs}\eta _s+i\bar{\xi}_iG_Q^{ij}\xi
_j\right\}, 
\]
and $N$ normalizes $Z$ to be one at vanishing sources. Use will be made of
the DeWitt compact notation \cite{Daemi}, in which a Latin letter let say 
$a,b,...$, for a field symbolizes the space-time coordinates of it as well as
all its internal quantum numbers. The same index in a source or a tensor
symbolizes the same set of variables of the kind of fields associated with
this specific index. For example $A_a=A_{\mu _a}^a(x_a),\ c_a=c^a(x_a)\ 
\text{and}\ \Psi ^i=\Psi ^i(x_i)$. Note that a particular convention has
been also employed in which the same letter is employed for the global index
(for all the coordinated and internal indices) and the one indicating the
internal coordinates in the explicit form of the quantities. Such a
procedure was useful for the calculations and we also think that will not
create confusion. As usual, repeated indices represent the corresponding
space time integrals and contracted, Lorentz, spinor or color components.

The following definitions also have been used in (\ref{funct2}):
\begin{eqnarray*}
S_{ijk...}^{\alpha } &\equiv &\left(\frac{\delta }{\delta \Phi _{i}}\frac{
\delta }{\delta \Phi _{j}}\frac{\delta }{\delta \Phi _{k}}...S^{\alpha
}\left(\Phi \right) \right) _{\Phi =0}\text{ \ for \ }\alpha = G,\ Gh,\ Q 
\text{ \ and }\Phi =A,\ \bar{c},\ c,\ \bar{\Psi}, \ \Psi \, \\
G_{\alpha }^{ij} &\equiv &-S_{\alpha,ij}^{-1}\text{\qquad \qquad \qquad
\qquad \qquad for \ } \alpha =G,\ Gh,\ Q,
\end{eqnarray*}
where $G$, $Gh$ and $Q$ mean the gluon, ghost and quark parts of the action
respectively.

There is only one main element determining the difference of the usual
Perturbative QCD expansion and the modified one considered in \cite
{1995,PRD,tesis}. It is related with the form of the gluon free propagator.
As proposed in \cite{1995} there is an additional term to the gluon free
propagator which is absent in the standard expansion. In \cite{tesis,PRD},
such a term was shown to be a consequence of a Wick expansion based in a
state constructed by acting on the usual vacuum with an exponential of pairs
of zero momentum gluon and ghost creation operators. The ghost propagator 
\cite{PRD,tesis} could remain unmodified if the parameter left after fixing
the form of the wave-function is taken as a real and positive one \cite
{PRD,tesis}. Such a selection will be employed here. The quark propagator is
not affected in any way by the gluon condensation as introduced in \cite
{1995,PRD,tesis}. However, as it will be commented in the summary, the
results of the present work led to the idea about a possible physical
relevance of introducing quark condensates along the similar lines as it was
done for the gluons. We consider the exploration of this possibility to be
exposed as one of the most interesting extensions of the work.

A technical point can be noticed for precision. The operator quantization
employed in \cite{PRD,tesis} takes the ghost fields as satisfying the
conjugation properties proper of the approach of Kugo and Ojima \cite{Kugo}.
However, at the level of the Feynman diagram expansion the difference with
the standard procedure is only a change of variables.

In accordance with above remarks and the results of \cite{1995,PRD,tesis}
the diagram technique defining the modified expansion has the following
basic propagators for the renormalized fields
\begin{eqnarray*}
G_G^{ab} &=&\delta ^{ab}g^{\mu _a\mu _b}\left[ \frac 1{k^2+i\varepsilon }
-iC\delta \left( k\right) \right], \\
G_{Gh}^{rs} &=&\delta ^{rs}\frac{\left( -i\right) }{k^2+i\varepsilon }, \\
G_Q^{ij} &=&\delta ^{ij}\frac{m+p^\mu \gamma _\mu }{\left(
m^2-p^2-i\varepsilon \right) }.
\end{eqnarray*}
and the same normal and counterterm vertices of the usual theory \cite{Muta}. 
The dimensional regularization and minimal
substraction (MS) is the renormalization procedure employed here.

Being defined the Feynman rules and the conventions to be used, below they
will be applied to the calculation of various quantities of interest.

\section{The mean value of $G^2$}

The calculation of the mean value of the gluon field intensity squared is
considered in this section. It extends the former evaluation done in \cite
{1995} within a simpler approximation. The result will be employed
afterwards for fixing the value of the condensation parameter $g^2C$ to be
the necessary one for furnishing the currently estimated value for $\langle
g^2G^2\rangle$. The calculation will be performed up to order $g^2$
including all the dependence on the condensate.

It is a helpful circumstance here, that the one loop renormalization
procedure of the modified expansion can be performed identically as in the
standard PQCD. This is a direct consequence of the fact that the 
$\delta$-functions kill the only existing integrals at the one loop level and
therefore all the corresponding counterterms in the dimensional
regularization are identical. This fact will be helpful in regularizing and
renormalizing the composite operator $\langle G^2\rangle$.

For the evaluation of $\langle G^2\rangle$ the following definition of this
quantity will be employed
\begin{equation}
\langle G^2 \rangle =\frac{\left\langle 0\right| S_g\left| 0\right\rangle }
{\int dx^D}=
\frac{\int D\left( A^r,\bar{c}^r,c^r,\bar{\Psi}^r,\Psi ^r\right) S_g\left[
A^r\right] \exp \left\{ i\int d^4x{\cal L}^{sources}\right\} }{\int dx^D\int
D\left( A^r,\bar{c}^r,c^r,\bar{\Psi}^r,\Psi ^r\right) \exp \left\{ i\int d^4x
{\cal L}\right\} },  \label{sg}
\end{equation}
where $S_g\left[ A^r\right]$ represents the gluon part of the action in the
absence of the gauge breaking term depending on the gauge parameter $\alpha $, 
evaluated at the renormalized gluon fields $A^r$, that is
\[
S_g\left[ A^r\right] =-\frac 14\int d^4xF_{\mu \nu }^{ra}\left( x\right)
F^{r\mu \nu,a}\left( x\right). 
\]

Therefore, $\langle G^2\rangle$ will be considered as defined by the mean
value of the gluon action divided by the space-time volume in dimensional
regularization.

After substituting in (\ref{sg}) the perturbative expression for the mean
value in terms of the Generating Functional (\ref{funct2}) it follows 
\begin{equation}
\left\langle 0\right| S_g\left| 0\right\rangle =\left\{ \left[ \frac{S_{ab}^g
}{2i^2}\frac{\delta ^2}{\delta J_a\delta J_b}+\frac{S_{abc}^g}{3!i^3}\frac{
\delta ^3}{\delta J_a\delta J_b\delta J_c}+\frac{S_{abcd}^g}{4!i^4}\frac{
\delta ^4}{\delta J_a\delta J_b\delta J_c\delta J_d}\right] Z\left[ J,\bar{
\eta},\eta,\bar{\xi},\xi \right] \right\} _{J,\bar{\eta},\eta,\bar{\xi}
,\xi =0}.  \label{des1}
\end{equation}
In the momentum representation the kernel $S_{ab}^g$ has the form
\[
S_{ab}^g=\left( 2\pi \right) ^4\delta \left( k_a+k_b\right) \delta
^{ab}\left[ -\left( g_{\mu _a\mu _b}k_a^2-k_{a,\mu _a}k_{a,\mu _a}\right)
\right], 
\]
and the $S_{abc}^g$ and $S_{abcd}^g$ ones are the usual three and four legs
gluon vertices \cite{Muta}.

Let us consider below the evaluation of $\langle G^2 \rangle$ up to order $g^2$. 
As any Feynman diagram can be expressed as a polynomial in the constant $C$, 
after associating a new diagrammatic line for this part of the free propagator,
all the Feynman diagrams can be divided in two groups. The first, let say $G_1$ 
including the diagrams not showing the new propagator line and another
group, $G_2$ having one or more, let us call them, ``condensate'' lines.
Such lines have associated to them a multiplicative Dirac delta function at
zero momentum.

\begin{figure}[h]
\begin{center}
\includegraphics[scale=0.29,angle=0]{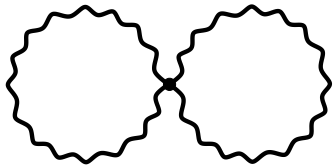}
\end{center}
\vspace{-0.7cm}\caption{Diagram contributing to $G_2$ terms in the gluon 
condensate
parameter.}
\label{G2}
\end{figure}

After writing out all the expressions of type $G_2$ (up to second order in
the renormalized coupling $g^2$) contributing to (\ref{des1}), it follows
that the only non-vanishing diagram is the one appearing in the Figure \ref
{G2} in which both gluon lines are substituted by ``condensate'' lines. The
remaining terms vanish thanks to the zero momentum evaluation produced by
the Dirac Delta functions introduced by the ``condensate'' lines and also
due to the vanishing of tadpoles in dimensional regularization. Also, it
should be underlined that the retention of the $i\epsilon $ factors defining
the Feynman propagator, up to the evaluation of the integrals was also
important in obtaining the zero result for some of the diagrams in $G_2$.

\vspace{0.5cm}

The contribution related with Figure \ref{G2} had been calculated before and
the result is \cite{1995,PRD}
\[
T_{G_2}=-\frac 14\frac{288g^2C^2}{\left( 2\pi \right) ^8}. 
\]

\begin{figure}[h]
\begin{center}
\includegraphics[scale=0.29,angle=0]{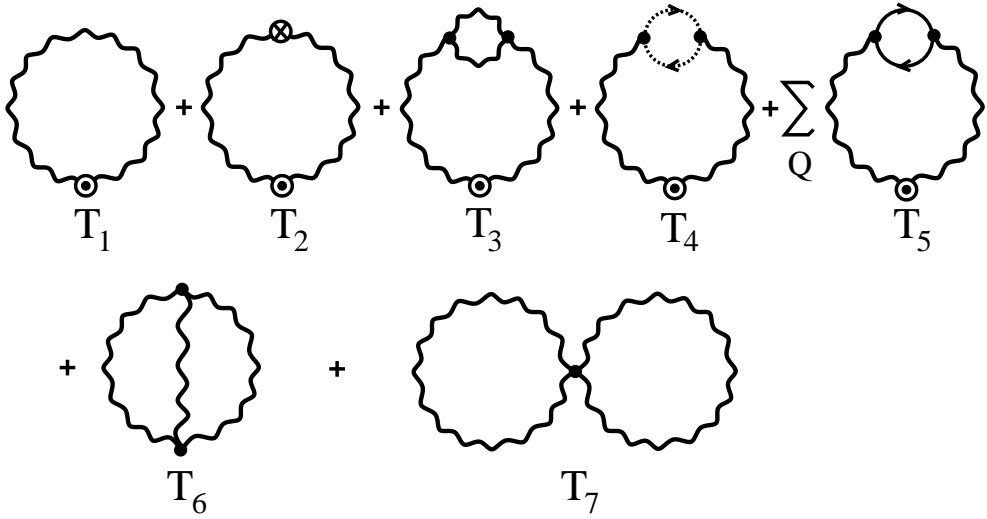}
\end{center}
\vspace{-0.7cm}
\caption{Diagram contributing to $G_1$ terms in the gluon condensate
parameter.}
\label{G3}
\end{figure}

Next, the diagrams in $G_1$ are the ones contributing to $G^2$ up to order 
$g^2$ in standard perturbative QCD. They are illustrated in Figure \ref{G3}.
The terms $T_6$ and $T_7$ represent the only non-vanishing and non-tadpole
like second order contributions. It can be noticed that the appearance of
two tadpoles in the term $T_7$ leads to the exact vanishing of this
contribution in dimensional regularization.

\vspace{0.7cm}

For the diagram $T_6$ corresponds the expression
\[
E_{T_6}=\frac 1{3!\ i^3}\int \frac{dp^D}{(2\pi )^Di}\frac{g^{\mu \nu }\delta
^{ab}}{p^2+i\varepsilon }\Pi _{\mu \nu }^{(g)\ ab}(p), 
\]
where $\Pi^{(g)}$ is the gluonic contribution to the polarization tensor in
second order, which has been explicitly calculated. After substituting 
the known result for $\Pi $ and performing the Wick
rotation, the previous expression turns out to be of the form 
\begin{equation}
E_{T_6}\backsim \int \frac{dq^D}{(2\pi )^D}\ (q^2)^{\frac D2-2}
\label{dimint}
\end{equation}

Then considering the formula \cite{narison}
\begin{eqnarray*}
\int \frac{dq^D}{(2\pi )^D}\frac 1{(q^2)^\alpha }\frac 1{(q-k)^{2\beta }} &=&
\frac{i(-1)^{-\alpha -\beta }(-k^2)^{-\alpha -\beta +n/2}}{(16\pi ^2)^{n/4}}
\frac{\Gamma (\alpha +\beta -n/2)}{\Gamma (\alpha )\Gamma (\beta )}B \left( 
\frac n2-\beta,\frac n2-\alpha \right), \\
B(n,m) &=&\frac{\Gamma (m)\Gamma (n)}{\Gamma (m+n)},
\end{eqnarray*}
for the case $\alpha +\beta =2-\frac D2$ and fixing $k_\mu =0$, it follows
the vanishing of diagram $T_6$ within the dimensional regularization for $D>2$.

It rest now to discuss the contributions of the tadpole like diagrams. It is
worth noticing at this point that they are quite similar in their form to
the one corresponding to the square of the electromagnetic field strength in
QED. In fact they differ from it only in some numerical color factors. An
important quantitative difference between the expressions for both theories
is related with magnitude of the coupling constant which is very much higher
in QCD. Let us consider the $T_1-T_5$ terms in Figure \ref{G3} embodying the
gluon, ghost and fermion contributions with the inclusion of the one loop
counterterms rendering finite the polarization operator at one loop 
\cite{Muta}. This expression turns to be of the form 
\begin{equation}
E_{T_1-T_5}\backsim \int \frac{dp^D}{(2\pi )^D}\ \frac{(p_\mu p_\nu
-p^2g_{\mu \nu })(g^{\mu \nu }-\frac{p^\mu p^\nu }{p^2+i\varepsilon })}{
p^2+i\varepsilon }\delta ^{aa}\left(1-\Pi (p^2)\right),  \label{T15}
\end{equation}
where $\Pi $ as usual is defined through the polarization tensor
\[
\Pi _{\mu \nu }^{ab}(p)=\delta ^{ab}(p_\mu p_\nu -p^2g_{\mu \nu })\Pi (p^2), 
\]
and it can be expressed as the superposition of the pure gluodynamical
contribution and the quark one as \cite{Muta} 
{\setlength\arraycolsep{0.6pt}\begin{eqnarray}
\Pi (p^2) &=&\Pi _G(p^2)+\Pi _Q(p^2) \nonumber \\
&=&\frac{g^2\mu ^{4-D}}{4\pi ^{D/2}}(-p^2)^{2-D/2}\Gamma \left( 2-\frac D2 \right)
B\left( \frac D2,\frac D2 \right) \times  \nonumber \\
&&\times \left( -\frac{C_G(3D-2)}{\frac D2-1}+\frac{8T_R}{B(\frac D2,\frac D2)}
\sum_Q\int_0^1dxx(1-x) \left( x(1-x)-
\frac{m_Q^2-i\varepsilon }{p^2} \right) ^{-(2-D/2)} \right)  \nonumber \\
&&+(Z_3-1),
\end{eqnarray}
}\noindent where $C_G=3$ and $T_R=1/2$ for $SU(3)$ and $\mu $ is the scale mass 
in the dimensional regularization.

Considering first the pure gluonic contributions associated $(1-\Pi _G(p^2))$
in (\ref{T15}), it follows that they reduce to a sum of dimensional
regularized integrals of the same form as (\ref{dimint}). Thus,
the second order gluodynamical contribution of the tadpole like
contributions vanish. Let us analyze now the quark contributions associated
to $\Pi _Q$. It can be noticed that for massless quarks, that is $m_Q=0$,
the $\Pi _Q$ has the same analytical dependence of $p^2$ as the gluonic part 
$\Pi _G$ and all the quark contributions vanish after the regularization is
employed. However, for massive quarks the result for the second order
dimensionally regularized contribution along the same lines employed for the
previous terms gives rise to a non-vanishing contribution of the following
form after the substraction of the divergent pole terms at $D=4$. The
expression for the regularized term before substraction is
\[
g^2\mu ^{4-D}E_{T_6}=\left( \sum_Q(m_Q^2)^{D-2}\ \right) \frac{64\sqrt{\pi}
g^4\mu ^{8-2D}T_R(D-1)\Gamma (2-D)\Gamma (4-D)}{2^{7-2D}(4\pi )^D\Gamma (
\frac 92-D)}. 
\]

This expression has simple as well as double poles which after substracted
(in the $D=4$ limit following the rules of the MS scheme for composite
operators \cite{collins}) give a finite correction to $\langle g^2G^2\rangle$ 
for the gluon condensate parameter. It can be evaluated and the numerical
expression becomes 
\[
\left( g^2\mu ^{4-D}E_{T_6}\right) ^{fin}=0.00064T_Rg^4\left( \sum_Q m_Q^4 
\left[ 55.32-37.4\ \ln \left( \frac{\mu ^2}{m_Q^2} \right)+6.0\ 
\ln^2 \left( \frac{\mu ^2}{m_Q^2} \right)
\right] \right). 
\]

However, it should be noticed that this finite expression has been obtained
for the second order correction of the polarization operator. The summation
of the infinite ladder of self-energy insertions leading to the one loop
propagator can be suspected to produce a noticeable modification of this
result. After also considering the high indefiniteness in the quark current
masses, and the technical difficulty we had in evaluating the dimensionally
regularized expression for the above mentioned self-energy insertions, we
decided to employ for the further consideration here the purely
gluodynamical contribution. In support of this way of proceeding there is
also another circumstance. It is needed to notice that a refined version of
the definition of the composite operator $G^2(x)$ could be suspected to
exist in such a way that the terms associated to a tadpole diagram of the
exact propagator vanish. This property would assure the exact vanishing of
the $\langle F^2\rangle$ in QED a property that to our knowledge, is expected. 
Because, at one loop, both QED and QCD tadpole expressions are basically the 
same, the above written quark contribution would not really appear. The existence
of the mentioned definition, opens the possibility for the second order
gluodynamical contribution to be exact one for QCD in this approximation.
However, we shift the discussion of this question to further analysis to be
considered elsewhere and simply adopt the gluodynamical result.

Therefore, the total second contribution to the mean value of $G^2$ in
gluodynamics is defined by the only non-vanishing term
\[
\langle G^2\rangle =\frac{288g^2C^2}{\left( 2\pi \right) ^8}, 
\]
coming from the condensate contributions arising from the diagram $T_6$
which is defined by the four gluon interactions. Finally, let us
determine in a phenomenological way a value for the constant $g^2C$ by
making use of a currently estimated value for $\langle g^2G^2\rangle $:
\[
\langle g^2G^2\rangle \cong 0.5\left( GeV/c^2\right) ^4. 
\]
Therefore, the parameter $g^2C$ takes the value:
\begin{equation}
g^{2}C=64.9394\ \left(GeV/c^{2}\right) ^{2}.  \label{g2}
\end{equation}

This fixing of the condensation parameter allows to investigate the physical
predictions of the modified expansion in the next section.

\section{Constituent Quarks masses}

\begin{figure}[h]
\begin{center}
\includegraphics[scale=0.20,angle=0]{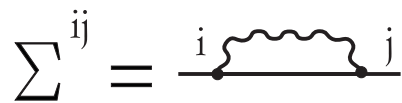}
\end{center}
\vspace{-0.7cm}\caption{Diagram for $\Sigma\left(p\right)$ in one loop.}
\label{autoQ}
\end{figure}

Let us consider now the effect on the current quark masses produced by the
condensate. The full inverse propagator for the quarks in terms of the
corresponding mass operator has the general expression
\begin{equation}
G_{Q}^{-1}\left(p\right) =\left(m_{Q}-p^{\mu }\gamma _{\mu }-\Sigma
\left(p\right) \right),  \label{prop1}
\end{equation}
where the color index identity matrix has been omitted and 
$\Sigma\left(p\right)$ is the mass operator. Its expression is related with
the diagram in Figure \ref{autoQ}.

The form for $\Sigma \left( p\right)$ up to order $g^2$, within the modified
Feynman rules can be written as
\begin{equation}
\Sigma \left( p\right) =g^2C_F\int \frac{d^4k}{\left( 2\pi \right) ^4i}\frac{
\gamma _\mu \left( m_Q+\left( p-k\right) ^\alpha \gamma _\alpha \right)
\gamma _\nu G_G^{\mu \nu }\left( k\right) }{\left( m_Q^2-\left( p-k\right)
^2\right) },  \label{auto}
\end{equation}
in which the color factor $C_F$ for the $SU\left( N\right)$ group is $C_F=
\frac{N^2-1}{2N}$ which for the case of interest here $SU\left( 3\right)$,
reduces to $C_F=\frac 43$ \cite{Muta}.

Taking into account the modified gluon propagator and the standard relations
for the $\gamma$ matrices (which here are adopted in the same way as in
Ref. \cite{Muta}) the one loop mass operator expression is simplified to be
\[
\Sigma \left( p\right) =g^2C_F\int \frac{d^4k}{\left( 2\pi \right) ^4i}\frac{
2\left( 2m_Q-\left( p-k\right) ^\alpha \gamma _\alpha \right) }{\left(
m_Q^2-\left( p-k\right) ^2\right) }\left( \frac 1{k^2}-iC\delta \left(
k\right) \right). 
\]
Let us disregard for the moment the term not involving the condensate
parameter $C$ in order to first study the sole influence on the quark masses
of the condensation effect. Then, the expression for $\Sigma $ reduces to
\[
\Sigma \left( p\right) =-\frac{g^2C_FC}{\left( 2\pi \right) ^4}\frac{2\left(
2m_Q-p^\alpha \gamma _\alpha \right) }{\left( m_Q^2-p^2\right) }. 
\]
Therefore, the inverse propagator for the quark takes the simple form
\begin{equation}
G_Q^{-1}\left( p\right) =m_Q\left( 1+2\frac{M^2}{\left( m_Q^2-p^2\right) }
\right) -p^\mu \gamma _\mu \left( 1+\frac{M^2}{\left( m_Q^2-p^2\right) }
\right),  \label{prop2}
\end{equation}
where the constant $M$ is defined by 
\[
M^2=\frac{2g^2C_FC}{\left( 2\pi \right) ^4}=0.1111\left( GeV/c^2\right) ^2, 
\]
in which the numerical value has been obtained by using (\ref{g2}).

The zeros of the determinant associated to the inverse propagator (\ref
{prop2}) then allows to determine the effects of the condensate (reflected
by the parameter $M$), on the effective mass of the quarks. Let below 
$s,s=1...6$ an index indicating each kind of quark characterized by its
particular current mass $m_{Q_s}$.

\begin{figure}[h]
\begin{center}
\includegraphics[scale=0.44,angle=-90]{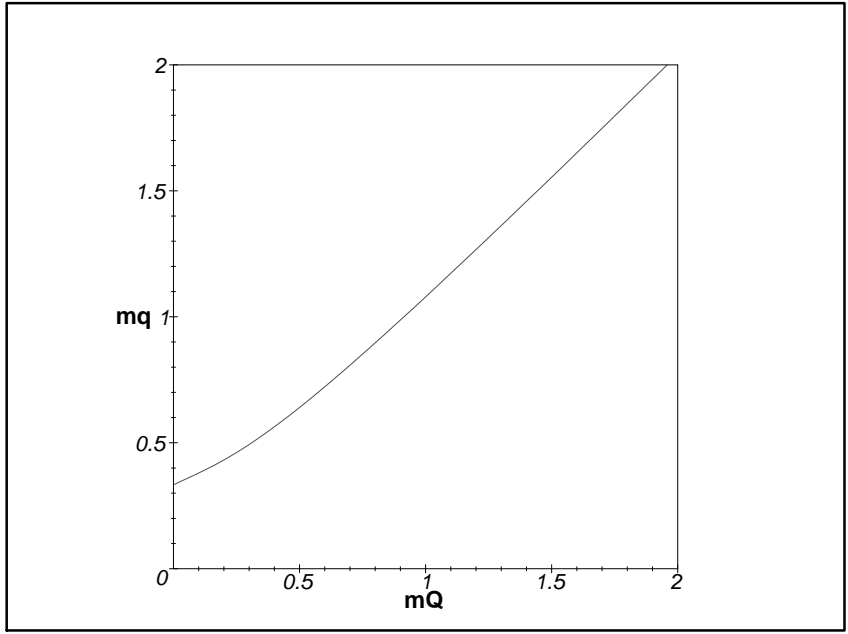}
\end{center}
\vspace{-0.7cm} \caption{Real solution for the quark mass as a
function of the Lagrangian mass (masses in $GeV/c^2$).}
\label{Mass1}
\end{figure}

In order to find the dependence of the effective quark mass $m_{q_{s,i}}^2$
as a function of the current mass parameters $m_{Q_s}$ an analysis was done
for the zeros of the determinant of the inverse green function (\ref{prop2}).
It follows that there is only one solution having a squared mass being
positive for arbitrary values of the current quark masses. The existence of
this real and positive solution for the squared mass becomes possible thanks
to the real and positive character of $C$, as was shown previously 
\cite{PRD,tesis}. There is another interesting branch of the zeros of the
determinant (dispersion relations) showing vanishing values for the
effective mass in the light current mass limit. For the purely real solution
the value of the effective quark mass $m_{q_s}=f\left( m_{Q_s}\right)$ as a
function of $m_{Q_s}$ is shown in the Figure \ref{Mass1}. The graph is
plotted for the region $m_{Q_s}<2\ GeV/c^2$ which contains the current mass
values of the $u,\ d,\ s$ and $c$ quarks.

As it can be appreciated in the picture, the calculated effective quark
masses for light flavors ($u,\ d$ and $s$) are clearly predicting the values
of the quark masses being in use in the constituent quark models of the
baryons. That is, the light quarks get a weight of near one third of the
nucleon mass. In the table \ref{table1} the mass values obtained for each of
the quark flavors are shown \cite{Report}. From the global properties of
table \ref{table1} it can be observed that the main effect of the gluon
condensate seems to be the dressing the light quarks with a cloud gluons
having a total mass of one third of the nucleon mass. Then, the results
point in the direction of the mainly glue nature of the constituent quark
masses of the $u,\ d$ and $s$ quarks within many baryon resonance. These
results then support the idea that a modified perturbative expansion like
the one being considered, in which the effect of the gluon condensate have
been incorporated, would be able to predict with reasonably good
approximation properties of the low energy strong interactions. Similar
values for the constituent quarks masses have been obtained by different
methods in \cite{Steele1,Steele2}.

\begin{table}[h]
\caption{Quark mass values in presence of the condensate in units of $MeV/c^2$.}
\label{table1}
\begin{tabular}{||c||c||c||c||}
\hline\hline
$Quarks$ & $m_{Low}^{Exp}$ \footnote{Reported lower bound value for the
Lagrangian mass} & $m_{Up}^{Exp}$ \footnote{Reported upper bound value for
the Lagrangian mass} & $m_{Med}^{Theo}$ \footnote{Calculated mean value of
the constituent mass, for the lower and upper bound of the Lagrangian masses 
as determined only by the condensate contribution to the self-energy in the
one loop approximation} \\ \hline\hline
(u) & 1.5 & 5 & 334.944 \\ \hline\hline
(d) & 3 & 9 & 336.287 \\ \hline\hline
(s) & 60 & 170 & 388.191 \\ \hline\hline
(c) & 1100 & 1400 & 1315.241 \\ \hline\hline
(b) & 4100 & 4400 & 4269.572 \\ \hline\hline
(t) & 168600 & 179000 & 173800.48 \\ \hline\hline
\end{tabular}
\end{table}

We will now investigate the influence on the mass spectrum produced by
including the standard one loop contribution in the dispersion relations.
After substituting the known \cite{narison} quark one-loop self-energy
expression, the one loop inverse quark propagator can be written as
\begin{eqnarray}
G_Q^{-1}\left( p\right) &=&\left( m_Q-p^\mu \gamma _\mu -\Sigma \left(
p\right) \right) =A(p^2)\ m_Q-B(p^2)\ p^\mu \gamma _\mu  \label{inver} \\
&=&m_Q \left[1+\frac{4C_F\ g^2C}{(2\pi )^4}\frac 1{m_Q^2-p^2}-
\frac{C_F\ g^2(\mu )}{(4\pi )^2}\times \right.  \nonumber \\
&& \left. \left( -6+4\gamma -\ln(4\pi )+4\ \ln \left(\frac{m_Q^2-p^2}
{\mu ^2} \right) -4\ \frac{m_Q^2}{p^2}\ln \left( \frac{m_Q^2-p^2}{m_Q^2} \right) 
\right) \right]  \nonumber \\
&&-p^\mu \gamma _\mu \ \left[1+\frac{2C_F\ g^2C}{(2\pi )^4}\frac 1{
m_Q^2-p^2}+\frac{C_F\ g^2(\mu )}{(4\pi )^2}\times \right.  \nonumber \\
&&\left. \left(1-\gamma +\ln(4\pi )+\frac{m_Q^2}{p^2}-\ln \left( \frac{m_Q^2-p^2}
{\mu^2}\right) +(\frac{m_Q^2}{p^2})^2\ln \left(\frac{m_Q^2-p^2}{m_Q^2}\right)
\right)\right], \nonumber
\end{eqnarray}
where the dimensionless coupling constant $g^2$ has been substituted by its
running expression as a function of the scale parameter $\mu$ as 
\[
g(\mu )^2=\frac{4\pi \ 0.12}{1+\frac{7\ 0.24}{4\pi }\ln\left( \frac 
\mu {91.18} \right) }. 
\]
That is, the value of $\alpha^2 (\mu )=g^2(\mu )/(4\pi)$ is fixed to be 
$0.12$ at the scale of the $Z$ mass $\mu =m_Z=91.18\ GeV$. In order to
automatically fix the scale $\mu $ to the one associated to the magnitude of
the obtained solution for the mass, the value of $\mu $ was taken to satisfy 
$\mu ^2=p^2$. That is, to reproduce the same value of the solution for the
squared mass.

Before proceeding, it should be noticed that the inverse propagator, after
the introduction of the one loop terms, gets a branch cut at the real axes
for the considered complex variable $p^2$ for Re$(p^2)>m_Q^2$. However,
the $i\epsilon$ prescription selects one of the sides of the cut as giving
a well defined value of the inverse propagator at real $p^2$ values.

After multiplying (\ref{inver}) by $A(p^2)\ m_Q-B(p^2)\ p^\mu \gamma _\mu $
the modified mass shells can be most easily found by solving 
\[
D(q^2)=A^2(p^2)\ m_Q^2-B^2(p^2)\ p^2=0. 
\]

Concretely, we estimated the corrected values of the masses in Table \ref{table1} 
by searching for the peaks of the inverse for the absolute value of $D(q^2)$ as
a 2D-function of the real and imaginary parts of $p^2$. The new values are
reported in Table \ref{table1b}. As it can be observed the masses were not 
drastically changed. They all show a relatively small reduction in their 
magnitude, then indicating the consistency of the predicted constituent masses 
for the light quarks at the one loop level.

\begin{table}[h]
\caption{Quark mass values in presence of the condensate in units of 
$MeV/c^2$.}
\label{table1b}
\begin{tabular}{||c||c||c||c||}
\hline\hline
$Quarks$ & $m_{Low}^{Exp}$ \footnote{Reported lower bound value for the
Lagrangian mass} & $m_{Up}^{Exp}$ \footnote{Reported upper bound value for
the Lagrangian mass} & $m_{Med}^{Theo}$ \footnote{Calculated mean value of
the constituent mass, for the lower and upper bound of the Lagrangian masses
after the inclusion of the standard one loop correction for the self-energy.}
\\ \hline\hline
(u) & 1.5 & 5 & 309.15 \\ \hline\hline
(d) & 3 & 9 & 309.15 \\ \hline\hline
(s) & 60 & 170 & 354.00 \\ \hline\hline
(c) & 1100 & 1400 & 1343-200$i$ \\ \hline\hline
(b) & 4100 & 4400 & 4515-65$i$ \\ \hline\hline
(t) & 168600 & 179000 & 190000-1450$i$ \\ \hline\hline
\end{tabular}
\end{table}

An interesting outcome following from the inclusion of the standard one loop
terms, should be stressed. It consists, in that for the smallest $m_Q$
masses associated to the lower and higher bounds for the $u$ and $d$ quarks, 
$\frac 1{\mid D(q^2)\mid }$ although showing a definite peak at $0.30915\
GeV $ does not really diverge near this value. Therefore, the propagator is
not showing a definite pole associated to a propagating quark mode at this 
$p^2$ value. However, at the level of the Bethe-Salpeter
equation for bound states of quarks, it should be expected that $p^2$ at
this peak should behave as the non-relativistic value for the square of the
mass entering the kinetic energy expression $\frac{p^2}{2m_Q}$. Henceforth,
in spite of the presence of free propagating modes in the simplest
approximation, the inclusion of the quantum one loop correction makes the $u$
and $d$ modes non-propagating (confined) in the same way as it occurs in the
approach of Munczek and Nemirovsky \cite{Munczek,Burden}.

For greater values of $m_Q$ in the range presently estimated for the $s$
quark current mass ($0.060-0.170\ GeV$) in the neighborhood of the
non-diverging peak of $\frac 1{\mid D(q^2)\mid }$, this quantity starts
showing divergency points at complex values of $p^2$ which represent
damped but propagating quark modes. The just described structure of the
spectrum suggests a possible explanation for the drastic difference of
stability between the $u$ and $d$ composed hadrons and the resonances
including $s$, $c$, $b$ and $t$ quarks. The point is that the presence of poles 
at complex $p^2$ for the heavier quarks could lead to the existence of
disintegration channels for the higher mass resonances in which the $s$, $c$, 
$b$ and $t$ quarks can escape in a damped propagating mode to be after
hadronized by some separate mechanism. On another hand, as the $u$ and $d$
quarks do not show propagating states, that could remain in a perennial
confinement within an undisturbed nucleon. A further study of the
consistency of this explanation is expected to be done within the planned
extension of this work to investigate bound state equations for hadrons.

It can be concluded that the constituent quark masses evaluated from the
simplest approximation for the dispersion relation considered above are
stable under the inclusion of the standard one loop corrections to the
self-energy.

Next, in the Table \ref{table2}, the masses of the ground state resonance
selected each of them from one of the groups of baryons as classified in 
\cite{Report}, are shown in comparison with the values obtained for them by
a theoretical estimate consisting in adding the here calculated masses for
each of their known quark constituents. The quark masses of the Table 
\ref{table1} were employed for this evaluation. Also, in the Table \ref{table3},
the masses for some vector mesons reported also in \cite{Report} are shown
in conjunction with the result of the sum of the calculated masses for their
constituent quarks (Table \ref{table1}).

\begin{table}[h]
\caption{Experimental and Theoretical Baryonic Resonance Masses in units of 
$MeV/c^2$.}
\label{table2}
\begin{tabular}{||r||r||r||r||}
\hline\hline
Baryon & Exp.Val. & Th.Mean.Val. & Rel.Err. \\ \hline\hline
p(uud) \ \  & 938.27231 & 1006.175 & 7.24 \\ \hline\hline
n(udd) \ \  & 939.56563 & 1007.519 & 7.23 \\ \hline\hline
$\Lambda $(uds) \ \  & 1115.683 & 1059.422 & 5.04 \\ \hline\hline
$\Sigma ^{+}$(uus)\ \ \  & 1189.37 & 1058.078 & 11.04 \\ \hline\hline
$\Sigma ^{0}$(uds) \ \  & 1192.642 & 1059.422 & 11.17 \\ \hline\hline
$\Sigma ^{-}$(dds)\ \ \  & 1197.449 & 1060.766 & 11.41 \\ \hline\hline
$\Xi ^{0}$(uss)\ \ \  & 1314.9 & 1111.325 & 15.48 \\ \hline\hline
$\Xi ^{-}$(dss) \ \  & 1321.32 & 1112.669 & 15.79 \\ \hline\hline
$\Omega ^{-}$(sss)\ \ \  & 1642.45 & 1164.572 & 29.10 \\ \hline\hline
$\Lambda _{c}^{+}$(udc)\ \ \  & 2284.9 & 1986.472 & 13.07 \\ \hline\hline
$\Xi _{c}^{+}$(usc) \ \  & 2465.6 & 2038.375 & 17.33 \\ \hline\hline
$\Xi _{c}^{0}$(dsc)\ \ \  & 2470.3 & 2039.719 & 17.43 \\ \hline\hline
$\Omega _{c}^{0}$(ssc)\ \ \  & 2704 & 2091.622 & 22.65 \\ \hline\hline
$\Lambda _{b}^{0}$(udb)\ \ \  & 5624 & 4940.803 & 12.15 \\ \hline\hline
\end{tabular}
\end{table}
\begin{table}[h]
\caption{Experimental and Theoretical Masses for a group of Vector Mesons in
units of $MeV/c^2$.}
\label{table3}
\begin{tabular}{||r||r||r||r||}
\hline\hline
Meson & Exp.Val. & Th.Mean.Val. & Rel.Err. \\ \hline\hline
$\rho \left( \frac{u\overline{u}-d\overline{d}}{\sqrt{2}}\right) $ & 770.0 & 
671.231 & 12.83 \\ \hline\hline
$\varpi \left( \frac{u\overline{u}+d\overline{d}}{\sqrt{2}}\right) $ & 781.94
& 671.231 & 14.16 \\ \hline\hline
$\phi \left( s\overline{s}\right) $ \ \ \ \  & 1019.413 & 776.381 & 23.84 \\ 
\hline\hline
$J/\psi \left( 1S\right) \left( c\overline{c}\right) $ & 3096.88 & 2630.482
& 15.06 \\ \hline\hline
$Y\left( 1S\right) \left( b\overline{b}\right) $ & 9460.37 & 8539.144 & 9.74
\\ \hline\hline
\end{tabular}
\end{table}

As it can be appreciated the results obtained for the vector mesons and
baryon resonance are only estimates. However, the obtained values are in a
reasonable well correspondence with the reported experimental values. It can
be also noticed that the validity of the present calculations implies that
the binding energy contribution to the baryon rest masses should be small.
This property then could justify a non relativistic description of such
resonances.

\begin{figure}[h]
\begin{center}
\includegraphics[scale=0.4,angle=-90]{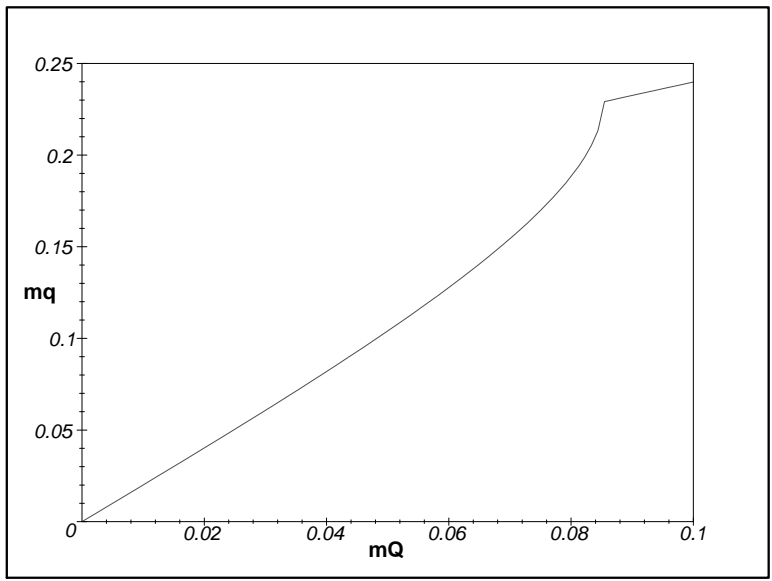}
\end{center}
\vspace{-0.7cm} \caption{A solution for the quark mass as a function of the 
Lagrangian mass, that satisfies $m_{q_s}\rightarrow0$ when $m_{Q_s}\rightarrow0$ 
(masses in units of $GeV/c^2$).}
\label{Mass2}
\end{figure}

As it was remarked before, another form of the dependence of the effective
quark mass with $m_{Q_s}$ is also following from the dispersion relations.
It, by contrast shows a vanishing effective mass when $m_{Q_s}\rightarrow 0$. 
The existence of this branch opens the possibility for the existence an
appropriate description of the light mesons ($\Pi $ and $K$) within the
proposed scheme. They could be associated to bound states of light quarks
propagating in such modes. The analysis of this question deserves a separate
treatment to be examined elsewhere. The graphic for the mentioned branch is
represented in Figure \ref{Mass2}, for the region of $m_{Q_s}<0.1\ GeV/c^2$, 
a point at which the change in the analytic dependence signals where the
solution gets an imaginary part when the current mass value $m_Q$ grows.

\section{ Tachyonic gluon Self-Energy}

\begin{figure}[h]
\begin{center}
\includegraphics[scale=0.38,angle=0]{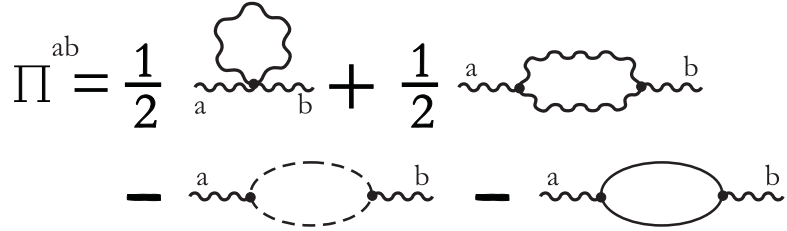}
\end{center}
\vspace{-0.7cm}\caption{Diagrams for $\Pi ^{ab}\left(p^{2}\right)$.}
\label{autoG}
\end{figure}

Let us consider now the evaluation of the gluon self-energy. As it is known
the exact gluon propagator can be written in the form
\begin{equation}
G_G^{ab}\left( p\right) =\frac{\delta ^{ab}}{p^2}\left( \frac{g_{\mu _a\mu
_b}-\frac{p_{\mu _a}p_{\mu _b}}{p^2}}{1+\Pi \left( p^2\right) }+\alpha \frac{
p_{\mu _a}p_{\mu _b}}{p^2}\right),  \label{Gprop1}
\end{equation}
where the function $\Pi$ is defined at the one loop level, by the expression
\begin{eqnarray*}
\Pi ^{ab}\left( p^2\right) &=&\Pi _G^{ab}+\Pi _T^{ab}+\Pi _{Gh}^{ab}+\Pi
_Q^{ab} \\
&=&\delta ^{ab}\left( p_{\mu _a}p_{\mu _b}-g_{\mu _a\mu _b}p^2\right) \Pi
\left( p^2\right),
\end{eqnarray*}
which is diagrammatically represented in Figure \ref{autoG}. 

Again, expanding in powers of the condensate parameter $C$, the total 
contribution can be decomposed in the standard one (associated to the ghost and 
the quark loop integrals) and the $C$ dependent ones $\Pi _G^{ab}$ and 
$\Pi _T^{ab}$ which can be readily calculated to be
\begin{eqnarray*}
\Pi _{G}^{ab} &=&\frac{g^{2}CN}{\left(2\pi \right) ^{4}}\delta
^{ab}\left(-5g_{\mu _{a}\mu _{b}}+2\frac{p_{\mu _{a}}p_{\mu _{b}}}{p^{2}}
\right), \\
\Pi _{T}^{ab} &=&\frac{3g^{2}CN}{\left( 2\pi \right) ^{4} }\delta
^{ab}g_{\mu _{a}\mu _{b}}.
\end{eqnarray*}

Summing up these particular two terms leads up to
\[
\Pi _C^{ab}\left( p^2\right) =\delta ^{ab}\left( p_{\mu _a}p_{\mu _b}-g_{\mu
_a\mu _b}p^2\right) \frac{2g^2CN}{\left( 2\pi \right) ^4p^2}. 
\]
Considering again our case the $SU\left( 3\right)$ group, that is $N=3$,
and disregarding in a first consideration the standard one loop
contribution, the function $\Pi $ reduces to
\begin{equation}
\Pi \left( p^2\right) =\frac{6g^2C}{\left( 2\pi \right) ^4}\frac 1{p^2},
\label{Gauto}
\end{equation}
which in turns leads out to the following condition for the poles of (\ref
{Gprop1})
\[
p^{2}-m_{G}^{2}=0, 
\]
where
\[
m_{G}^{2}=-\frac{6g^{2}C}{\left(2\pi \right) ^{4}}=-0.25\left(GeV/c^{2}
\right)^{2}. 
\]

Therefore, as the parameter $C$ has been defined before as real and positive
\cite{PRD,tesis}, it follows that the transverse gluon mass correction
becomes tachyonic. The ability of a tachyonic mass in producing improvements
in models for the inter quark potential has been recently argued in the
literature \cite{tachyon1,tachyon2}. The effect is linked with the
introduction of a linearly rising term as a first correction to the Coulomb
potential in the massless case \cite{tachyon2}. Therefore, the tachyonic
result arising here for the gluon mass appears to be of interest for a
derivation, within the present framework, of former successful
phenomenological bound state models for mesons \cite{sommer1,sommer2,El-Hady}. 
We are grateful to J.P. Vary and J. Qiu for the suggestion of this
possibility. It should be also underlined that the tachyonic character of
the gluon self-energy arising here, is fully compatible with the more
general result of Fukuda for the normal Green functions in QCD, when gluon
condensation is present \cite{fukuda}. This circumstance then suggests, that
the modified expansion under discussion could be a sort of perturbative
representation of the Fukuda non-perturbative treatment of the gluon
condensation in QCD. In a different manner, the approach of Celenza and
Shakin, and other authors can be expected to be related with a different
initial state also incorporating gluon condensation. It could be the point
that, at the end, both pictures can appear to be equivalent at the level of
exact results. This could be the case because the physical degrees of
freedom are not the gluons and quarks ones, but their colorless composite
combinations which physics is determined by the bound state dispersion
relations.

Finally in this section it deserves to be stressed that if a modification of
the ghost propagator is present, then a mass terms for the longitudinal
propagator could also appear and the polarization operator (gluon
self-energy) becomes non-transverse. This result means a violation of a Ward
identity representing the gauge invariance by the presence of ghost (and
fermion) condensates. Thus, it follows that the selection of the parameter
of the initial state in \cite{PRD,tesis} was the appropriate one for the
satisfaction of a gauge invariance condition at this stage. The physical
meaning of the alternative procedures of constructing the initial state
needs however for clarification.

\section{Summary}

The following results can be enumerated:

1) A calculation of the mean value of $G^2$ up to order $g^2$ in
gluodynamics, was performed. The result coincided with a previous one
obtained in a simpler approximation \cite{1995,PRD} and which signals that
quartic gluon self-interaction is relevant in fixing a non-vanishing gluon
condensation parameter already in the first order of the perturbation theory 
\cite{Zakharov}.

3) The one loop corrections for the quark masses under the influence of the
condensate were evaluated. When only the terms reflecting the effect of the
condensate are considered, the results for the constituent masses showed a
branch which is real as a function of all the values of the quark current
mass. Another solution which is real up to a critical value of the
Lagrangian masses, and having a vanishing proper mass in the limit of small
current masses is also obtained. The former branch predicts values of one
third of the nucleon mass for the masses of the light quarks. The effective
quark masses for all the quark flavors were evaluated as determined by their
known current mass in the mentioned approximation.

4) After the inclusion of the rest of the terms completing the full one loop
contribution to the quark self-energy, an interesting outcome arises. The $u$
and $d$ quarks, although showing high peak values of the propagator
components near the former values of the squared proper masses, become
confined. That is, their propagators do not show pole structures. On the
other hand the $s$, $c$, $b$ and $t$ quarks acquire poles but now not at
real values of $p^2$. Therefore, these free modes are damped. These
properties suggest a possibility for the explanation of the disparity of
stability properties in the nucleons and higher resonances. It would work as
follows, the absence of propagation solutions for the $u$ and $d$ quarks
could assure the unlimited life time of nucleons. Moreover, the damped
oscillations present for the rest of the quarks can create decay modes for
the bound states composed of them.

5) By mean of the obtained mass values, the ground state energies selected
within each of the groups hadron resonance reported in \cite{Report}, were
estimated through the simple addition of the masses of their known quark
constituents. It could be remarked that the obtained values are reasonable
ones in comparison with the data, after taking into account the high errors
in the experimentally determined Lagrangian masses. This fact supports the
hypothesis about the main glue nature of the constituent mass values of the
light quarks. The small contribution of the binding potential energy of such
resonance to their total mass is also indicated and thus suggests the
applicability of non relativistic approximations in their study.

6) The contribution of the condensate to the gluon self-energy of order $g^2$
was calculated. A tachyonic result for the gluon mass arises for the
parameter values employed in the construction of the initial state defined
in \cite{PRD,tesis}. This conclusion is in concordance with the analysis of
Fukuda for gluon condensation \cite{fukuda}. As it was remarked before,
there are recent arguments claiming the possibility of a tachyonic gluon
mass \cite{tachyon1,tachyon2,Hoyer3} and its role in improving the quark
antiquark potential in bound state models of mesons \cite{tachyon2}.

It seems useful to comment here some basic issues related with the modified
expansion being considered:

a) The Feynman expansion depending on the condensate parameter and taken for
the gauge parameter $\alpha =1$ corresponds with the Wick expansion around a
physical state of the interaction free QCD. In this somewhat limited way the
arbitrariness of the gauge parameter in the result for the gluon mass in 
\cite{1995} was solved in \cite{PRD,tesis}. However, in order to furnish a
fully consistent picture including renormalization, a physically equivalent
version of the expansion for arbitrary values of $\alpha $ should developed.
This question will be addressed in future works.

b) In \cite{PRD,tesis} the condensation parameter $C$ was defined as a real
and positive one. This result determined the tachyonic character for the
gluon mass found here. Further, the sign of $C$ also fixes the real values
of the order of the one third of the nucleon mass for the constituent quark
masses also calculated in the present work. The opposite sign produces
confined quark modes.

c) An aspect which is important to underline is that the selection of the
parameters defining the initial state in \cite{PRD,tesis} was designed also
in order to impose the absence of a modification for the free ghost
propagator. However, as it followed from the present analysis, this property
in turns, is related with the fact that a condensate modification of the
ghost propagator can produce a longitudinal contribution to the self-energy.
But, such a term should not exist for the transversality condition of the
polarization operator (a Ward identity) to be obeyed. Thus, its appearance
could break manifestly the gauge invariance. Therefore, the present work is
also given foundation to the non-modified ghost propagator choice considered
in \cite{1995,PRD,tesis}.

At the present stage we identify some important questions to be addressed in
the further extension of the work:

1) To explore the possibility of showing the exact gauge parameter
independence of the physical results within the modified expansion.

2) To address the general proof of the renormalizability in the expansion.

3) To search for a derivation of existing successful bounded state models
for the heavy quark mesons \cite{sommer1,sommer2,El-Hady}, as realized by
the a ladder approximation for the Bethe-Salpeter equation within the
proposed modified expansion. The presence of a tachyonic gluon propagator in
the approach (which is argued to have the effect of introducing a linearly
rising component in the inter-quark potential \cite{tachyon2}) and the
obtained constituent values for the light quark masses already support such
a possibility.

4) Improve the study of the effective action as a functional of the
condensate parameter done in \cite{1995} by also including the gauge field
in order to search for a variant of the leading logarithm model useful for
the investigation of field configurations associated to inter-quark strings,
nucleon-nucleon potentials, etc.

5)  It can be underlined that the considered expansion seems that could
furnish a general way for the perturbative description of pair condensates
in Many Body theory. An important example coming to the mind is the BCS
state of the usual superconductivity. Pair states being more closely
connected with QCD were discussed in \cite{Mishra,Diakonov}. In addition, in 
\cite{Diakonov} it was advanced before the idea of using diquark condensates
as generating an alternative for the Higgs mechanism. 

Finally, we would like to conjecture about a possibility suggested by the
results of this work. It is related with the question about whether the mass
spectrum of the whole three generations of fundamental fermions could be
predicted by a slight generalization of the modified perturbation expansion
under consideration. In this sense, the given arguments led us to the idea
that (after the introduction of quark condensates along the same lines as it
was done for gluon ones in \cite{1995,PRD,tesis}) the received perturbation
expansion can have the chance to predict both, the Lagrangian mass and the
constituent quark mass spectra of the three families of fundamental
fermions. The fermion condensates as described in the proposed perturbative
way, would have the role of producing the Lagrangian quark masses, through
the chiral symmetry breaking. The gluonic condensates, in one hand, and as
illustrated here, could be responsible of generating states of large
constituent mass for the low mass quarks states. In another hand, it seems
feasible that the higher order radiative corrections (including color
interactions with the condensate as well as chiral symmetry corrections)
could also determine the mass spectra for leptons and neutrinos. The smaller
scale for the masses of these particles could be produced by the lack of
lower order either color or electric-weak interaction terms in their
self-energy. Therefore, the possibility that the Lagrangian mass spectrum of
the three generations of the fundamental fermions could be predicted by a
modified perturbation expansion of the sort being proposed is suggested.
Work directed to investigate this possibility will be considered elsewhere.
The modified expansion under study could be a useful technique of
implementing this idea.

\begin{acknowledgments}

The authors wish to deeply acknowledge the helpful advice and comments of
Profess. J. Pestieau, J. Alfaro, M. Lowe, C. Friedli, M. Hirsch and A.
Gonz\'{a}lez. One of the authors (A. C. M.) also would like to express his
gratitude to the Abdus Salam ICTP, and in particular to its
Associate-Membership Program for the whole general support. The support of
the International Institute for Theoretical and Applied (IITAP) (UNESCO and
Iowa State University) and the Christopher Reynolds Foundation during the
completion of this work at IITAP is also greatly acknowledged. The
suggestion of the possibility of applying the procedure for describing bound
state model for mesons is acknowledged from Dr. J. P. Vary and Dr. J. Qiu.
Finally, we are grateful to K. Torn\'{e}s for the help in the preparation of
the manuscript.

\end{acknowledgments}


\begin{thebibliography}{99}

\bibitem{Savvidy1}  G. K. Savvidy, {\it Phys. Lett. B}71, 133 (1977).

\bibitem{Savvidy2}  I. A. Batalin, S. G. Matinyan, and G.K. Savvidy, {\it 
Sov. J. Nucl. Phys. }26, 214 (1977).

\bibitem{Savvidy3}  S. G. Matinyan, and G. K. Savvidy, {\it Sov. J. Nucl.
Phys. }25, 118 (1977); {\it Nucl. Phys. B}134, 539 (1978).

\bibitem{Reuter}  A. Cabo, O.K. Kalashnikov and A.E. Shabad, {\it Nucl.
Phys. B}185, 473 (1981).

\bibitem{ShuryakTex}  E. U. Shuryak, {\it The QCD Vacuum, Hadrons and the
Superdense Matter}, World Scientific, Singapore, 1988.

\bibitem{Shuryak2}  T. Sch\"{a}fer and E. V. Shuryak, {\it Rev. Mod. Phys.}
70, 323 (1998).

\bibitem{Gell}  H. Fritzsch, M. Gell-Mann, and H. Leutwyler, {\it Phys.
Lett. B}47, 365 (1973).

\bibitem{Gross}  D. J. Gross and F. Wilczec, {\it Phys. Rev. D}8, 3497
(1973).

\bibitem{Weinberg1}  S. Weinberg, {\it Phys. Rev. Lett.} 31, 494 (1973).

\bibitem{Weinberg2}  S. Weinberg, {\it The Problem of Mass}, Trans. New York
Acad. Sci. 38: 185-201, 1977; HUTP-77/A057, (1977).

\bibitem{Weinberg3}  S. Weinberg, {\it Phys. Rev. Lett. }64, 1181 (1990).

\bibitem{Weinberg4}  S. Weinberg, {\it Phys. Rev. Lett.} 67, 3473 (1991).

\bibitem{1995}  A. Cabo, S. Pe\~{n}aranda and R. Martinez, {\it Mod. Phys.
Lett. A}10, 2413 (1995).

\bibitem{PRD}  M. Rigol and A. Cabo, {\it Phys. Rev. D}62, 074018 (2000);
hep-th/9909057 (1999).

\bibitem{tesis}  M. Rigol, {\it Acerca de un Estado de Vac\'{\i}o
Alternativo para la QCD Perturbativa}, Graduate Dissertation Thesis,
Instituto Superior de Ciencias y Tecnolog\'{\i}a Nucleares, La Habana, Cuba,
1999, hep-th/0109012 (2001).

\bibitem{Hoyer1}  P. Hoyer, NORDITA - 96/63 P (1996), hep-ph/9610270 (1996).

\bibitem{Hoyer2}  P. Hoyer, NORDITA - 97/44 P (1997), hep-ph/9709444 (1997).

\bibitem{Hoyer3}  P. Hoyer and J. Rathsman, {\it JHEP} 05, 020 (2001).

\bibitem{Zakharov} M. A. Shifman, A. I. Vainshtein and V. I. Zakharov,
{\it Nucl. Phys. B}147, 385 (1979); {\it B}147, 448 (1979); {\it B} 147, 519  
(1979).

\bibitem{Kugo}  T. Kugo and I. Ojima, {\it Prog. Theor. Phys. Suppl.} 66, 1 
(1979).

\bibitem{Report}  C. Caso, et.al., {\it Review of Particle Physics, The
European Physical Journal C} 3, 1 (1998).

\bibitem{Munczek}  H. J. Munczek and A. M. Nemirovsky, {\it Phys. Rev. D}28,
181 (1983).

\bibitem{Burden}  C. J. Burden, C. D. Roberts and A. G. Williams, {\it Phys.
Lett. B}285, 347 (1992).

\bibitem{celenza}  L. S. Celenza and C. M. Shakin, {\it Phys. Rev. D}34, 
1591 (1986).

\bibitem{pavel}  H. P. Pavel, D. Blaschke, V. N. Pervushin and G. R\"{o}pke, 
{\it Int. J. Mod. Phys. A}14, 205 (1999).

\bibitem{tachyon1}  K. G. Chetyrkin, S. Narison and V. I. Zakharov, {\it Nucl.
Phys. B}550, 353 (1999).

\bibitem{tachyon2}  S. J. Huber, M. Reuter and M. G. Schmidt, 
{\it Phys. Lett. B}462, 158 (1999).

\bibitem{Muta}  T. Muta, {\it Foundations of Quantum Chromodynamics}, World
Scientific Lectures Notes in Physics - Vol. 5, Singapore, 1987.

\bibitem{Faddeev}  L. D. Faddeev and A. A. Slanov, {\it Gauge Fields.
Introduction to Quantum Theory}, Benjamin Cummings Publishing, Reading,
Mass., 1980.

\bibitem{Daemi}  S. Randjbar-Daemi, {\it Course in Quantum Field Theory},
Lectures Notes ICTP Diploma Programme, ICTP, Trieste 1994.

\bibitem{narison}  S. Narison,{\it Phys. Rep. }84, 263 (1982).

\bibitem{collins}  J. C. Collins, {\it Renormalization}, Cambridge University
Press, Cambridge, 1987.

\bibitem{Steele1}  V. Elias, T. G. Steele and M. Scadron, {\it Phys. Rev. D}38, 
1584 (1988).

\bibitem{Steele2}  V. Elias and T.G. Steele, {\it Phys. Lett. B}212, 88 
(1988).

\bibitem{sommer1}  A. J. Sommerer, J. R. Spence and J. P. Vary, 
{\it Phys. Rev. C}49, 513 (1994).

\bibitem{sommer2}  A. J. Sommerer, A. Abd El-Hady, J. R. Spence and J. P. Vary, 
{\it Phys. Lett. B}348, 277 (1995).

\bibitem{El-Hady}  A. Abd El-Hady, A. A. K. Lodhi and J. P. Vary, {\it Phys.
Rev. D}59, 094001-1 (1999).

\bibitem{fukuda}  R. Fukuda, {\it Phys. Rev. D}21, 485 (1980).

\bibitem{Mishra}  A. Mishra, H. Mishra, V. Sheel, S. P. Misra and P. K. Panda, 
{\it Int. J. Mod. Phys. E}5, 93 (1996), hep-ph/9404255 (1994).

\bibitem{Diakonov}  D. Diakonov,{\it \ Phys. Lett. B}373, 147 (1996),
hep-ph/9512385 (1995).

\end{thebibliography}
\end{document}